
\input phyzzx.tex
\pubnum{ROM2F-94-35}

\def\b{\beta}

\def\eps{\epsilon}
\def\vareps{\varepsilon}

\def\m{\mu}
\def\n{\nu}

\def\s{\sigma}
\def\t{\tau}

\def\mn{\mu\nu}
\def\mnr{\mu\nu\rho}

\def\wh{\widehat}
\def\wt{\widetilde}
\def\de{\partial}

\def\mezzo{{1 \over 2}}
\def\unoar{1-({a \over r})^4}

\def\ap{\alpha^\prime}
\def\unita{{1 \kern-.30em 1}}
\def\fey{{\big / \kern-.80em D}}
\def\complex{{\kern .1em {\raise .47ex \hbox {$\scriptscriptstyle |$}}
\kern -.4em {\rm C}}}
\def\zet{{Z \kern-.45em Z}}
\def\real{{\vrule height 1.6ex width 0.05em depth 0ex
\kern -0.06em {\rm R}}}
\def\rational{{\kern .1em {\raise .47ex \hbox{$\scripscriptstyle |$}}
\kern -.35em {\rm Q}}}
\titlepage
\title{ALE Instantons in String Theory\foot{Talk given by M.Bianchi
at the
Seventh Marcel Grossmann Meeting, Stanford July 24-30, 1994.}}
\author{M.~Bianchi, F.~Fucito, G.C.~Rossi}
\address{Dipartimento di Fisica, Universit\`a di Roma II ``Tor
Vergata''
I.N.F.N. Sezione di Roma II ``Tor Vergata'',00133 Roma, Italy}
\andauthor{ M.~Martellini}
\address{Dipartimento di Fisica, Universit\`a di Milano, 20133
Milano, Italy
Sezione I.N.F.N. dell'Universit\`a di Pavia, 27100 Pavia, Italy}
\abstract{In this talk we will describe a string solution which contains
a self-dual (instantonic) metric and study its properties.}

Asymptotically Locally Euclidean (ALE) Instantons have played a
fundamental role in Euclidean Quantum (Super)-Gravity [1]. Their
contribution to the formation of a gravitino condensate may trigger
dynamical supersymmetry breaking [2, 3].
The purpose of this talk is twofold.
The first is to promote ALE instantons to fullfledge
solutions to the heterotic string equations of motion.
The second is to study the geometrical properties of the simplest among them,
an Eguchi-Hanson instanton coupled to a gauge
instanton through the ``standard embedding".
This investigation is propaedeutical to a saddle-point evaluation of
instanton dominated Green functions in the low-energy effective
supergravity arising from the heterotic string compactified to D=4 [4].
Non trivial supersymmetric solutions of the lowest order (in $\ap$) equations
of motion may be found setting to zero the fermion fields together with their
supersymmetric variations. A supersymmetric ansatz for the solution
is given by [5]:
$$
F_{\mu\nu} = \wt F_{\mn} \quad
H_{\mnr}= \sqrt{G} \vareps_{\mnr}{}^\sigma\de_\sigma\phi \quad
G_{\mn} =  e^{2\phi} \wh g_{\mn} \quad
\eqn\ansatz
$$
where $F_{\mn}$ is the field-strength of the gauge fields $A_\m$, $H_{\mnr}$
is the (modified) field-strength of the antisymmetric tensor $B_{\mn}$
and $\wh g_{\mn}$ is a self-dual metric. The generalized spin-connection
with torsion $\Omega_{\m+}^{ab}$ deriving from \ansatz~ is self-dual.
\ansatz~ must be supplemented with the Bianchi identity:
$dH=\ap \{ trR(\Omega_{-})\wedge R(\Omega_{-})-trF(A)\wedge F(A) \}$.
To simplify matters it is convenient to impose the ``standard embedding''
of the gauge connection in the $SU(2)$ spin group: $A=\Omega_-$.
There are two options. The first has been considered in [5] and leads to
conformally flat axionic instantons.
We would like to concentrate on the other, \ie~
$ A^i_\mu ={1 \over 2} \eta^i_{ab} \Omega_{\m-}^{ab}
= {1 \over 2} \eta^i_{ab} \wh\omega_{\m}^{ab}$, whose consistency requires
a constant dilaton and a vanishing torsion [4].
In this case, \ansatz~is completely specified by the choice of a self-dual
metric $\wh g_{\mn}$.

An interesting class of self-dual metrics is given by the
Gibbons-\break
Hawking multi-center (GHMC) ansatz [1]:
$$
ds^2 = V^{-1}({\vec x}) (d\t + {\vec\omega}\cdot d {\vec x})^2 +
V({\vec x})d{\vec x}\cdot d {\vec x}
\eqn\multicenter
$$
with ${\vec \nabla}V={\vec \nabla}\times{\vec\omega}$ and
$V({\vec x}) = \eps + 2m \sum_{i=1}^{ k+1} {1\over\mid {\vec x}-
{\vec x}_i\mid}$.
The choice $\eps=0, m=\mezzo$ corresponds to ALE metrics.
ALE manifolds are smooth resolutions of singular varieties
in $\complex^3$ and are completly classified in terms
of the kleinian subgroups $\Gamma$ of $SU(2)$ [1].
ALE instantons are non-compact Ricci-flat hyperk\"ahler manifolds of
$SU(2)/\Gamma$ holonomy and deserve to be considered
as non-compact Calabi-Yau manifold of complex dimension two [6].

The non-linear $\s$-model describing the propagation of the
heterotic string on ALE instantons with the standard embedding is left-right
symmetric and admits $N=(4,4)$ supersymmetry [7].
The corresponding $\b$-functions vanish to all orders [7] thus
the lowest order ALE solutions receive no radiative corrections in $\ap$.
Indeed, at the singular point of the moduli space where ALE instantons
coincide with algebraic varieties, string propagation is governed by a
$\complex^2/\Gamma$ orbifold conformal field theory [6].

We now turn to describe some geometrical properties of the
heterotic solution based on the EH (Eguchi-Hanson) instanton [1]:
$$
ds^2= \wh g_{\mn} dx^\mu dx^\nu=
({r\over u})^2 dr^2+r^2 (\sigma_x^2+\sigma^2_y)+ u^2 \sigma_z^2
\eqn\ehmetric
$$
To remove the bolt singularity at $r=a$, we change the radial variable to
$u=r \sqrt{\unoar}$ and identify antipodal points. The EH instanton
has an $S^3/Z_2$ boundary and admits
an $SU(2)_R \otimes U(1)_L$ isometry group. Its Euler characteristic $\chi$
is two and its Hirzebruch signature $\tau= b^+_2-b^-_2$ is one.
Exploiting the global right-handed supersymmetry generated by the
covariantly constant spinor $\bar\eps$,
the two (left-handed) gravitino zero-modes on EH may be
expressed in terms of the closed self-dual two-form [3].
These zero-modes are generated by the broken global left-handed
supersymmetry with parameter $\eta=u\eta_0$ ($\eta_0=i\s_0\bar\eps$):
$\psi_\mu=D_\mu \eta-{1\over 4}\s_\m \> \fey\eta$ [4].
The three zero-modes of the metric can be obtained
performing a further supersymmetry transformation:
$h_{\mn}= \nabla_\m \xi_\n + \nabla_\n \xi_\m - \mezzo (\nabla\cdot\xi)g_{\mn}$
with $\xi_\m = \eta\s_\m\bar\eps$.
A suitable interpretation of the infinitesimal diffeomorphisms $\xi^\m$
allows to relate them to the lack of invariance of the EH metric
under dilatation and the two rotations in the coset $SU(2)_L/U(1)_L$.
No dilatino zero-modes are expected on EH since the index of the Dirac
operator (for gauge singlets) is zero [1].
However the presence of ``charged" spinor zero-modes is guaranteed by
a non-vanishing index of the Dirac operator coupled to the gauge bundle $V$.
Embedding the instantonic $SU(2)$ in the ``hidden" $E(8)$ only the gauginos
will be affected.  With respect to the subgroup, $SU(2)\otimes E(7)$,
the adjoint of $E(8)$ breaks as: $\underline{248}=(\underline3,\underline1)
\oplus(\underline2,\underline{56})\oplus (\underline1,\underline{133})$.
The general formula for the index of the Dirac operator, $\fey_V$, coupled
to a vector bundle $V$ may be found in [1]. Performing the necessary
manipulations, we find:
$ind(\fey_{\underline2},M,\de M)=-1=\tau$
and $ind(\fey_{\underline3},M,\de M)=-6$.
Since $\fey_V$ has no right-handed zero-modes, one (normalizable)
left-handed gaugino zero-mode is expected for each of the 56 doublets
and six for the triplet.
The explicit expression of the gaugino zero-modes in the doublet can be easily
found by exploiting the explicit form of the normalizable harmonic self-dual
two-form on the EH manifold.
The six triplet zero-modes may be found by explicitly solving the Dirac
equation. Their expression has no clear geometrical meaning [4].
Thanks to the global right-handed supersymmetry, each zero-mode of the
gaugino generates two zero-modes for the gauge fields.

As is well known the bosonic zero-modes are related to the collective
coordinate of the heterotic EH background.
Thanks to the cancellation of the non-zero-mode functional determinants,
the evaluation of the relevant correlation
functions reduces to an integration of zero-modes of the fermi fields
over the finite-dimensional moduli space of the heterotic EH instanton.
In [4] the saddle-point approximation is performed and the consistency
of the result with Ward identities of a properly defined global supersymmetry
is checked. The interpretation in terms of condensates as well as
the relation to topological amplitudes in string theory and
to topological field theories [8] is under investigation [4].
\section{References}
\medskip
\item{1.}
T.~Eguchi, P.B.~Gilkey and A.J.~Hanson, Phys.Rep. {\bf 66} (1980) 213.
\item{2.}
E.~Witten, Nucl.Phys. {\bf B185} (1981) 513.
\item{3.}
K.~Konishi, N.~Magnoli and H.~Panagopoulos,
Nucl.Phys. {\bf B309} (1988) 201; ibid. {\bf B323} (1989) 441;
S.W.~Hawking, C.N.~Pope, Nucl.Phys. {\bf B146} (1978) 381.
\item{4.}
M.~Bianchi, F.~Fucito, M.~Martellini, G.C.~Rossi,
{\it ALE Instantons in Effective String Theory}, preprint ROM2F-17-94
and in preparation.
\item{5.}
S.-J.~Rey, Phys.Rev. {\bf D43} (1991) 526; C.~Callan, J.~Harvey and
A.~Strominger, Nucl.Phys {\bf B359}(1991) 611; ibid. {\bf B367} (1991)
60; R.R.~Khuri, Nucl.Phys. {\bf B387} (1992) 315.
\item{6.}
D.~Anselmi,
M.~Billo', P.~Fre', L.~Girardello, A.~Zaffaroni, Int. Jou. Mod. Phys. {\bf
A9}(1994)3007.
\item{7.}
N.J.~Hitchin, A.~Karlhede, U.~Lindstr\"om and
M.~Ro\v cek, Comm.Math.Phys. {\bf 108} (1987) 535;
L.~Alvarez-Gaum\'e and D.Z.~Freedman, Phys. Rev.
{\bf D15} (1980) 846; Comm. Math. Phys. {\bf 80} (1981) 443;
L.~Alvarez-Gaum\'e, Nucl.Phys. {\bf 184} (1981) 180;
L.~Alvarez-Gaum\'e and P.~Ginsparg, Comm. Math. Phys. {\bf 102}
(1985) 311; C.~Hull, Nucl.Phys. {\bf 260} (1985) 182.
\item{8.}
I.~Antoniadis, E.~Gava, K.~Narain, T.~Taylor,
{\it Topological Amplitudes in String Theory}, preprint IC/93/202;
E.~Witten, Comm. Math. Phys. {\bf117} (1988) 353.
\end